\documentclass[aps,prl,twocolumn,groupedaddress,amsmath,amssymb]{revtex4}
\usepackage[dvips]{graphicx}
\DeclareGraphicsExtensions{.eps,.ps}
\usepackage{bm}      % RevTex4 bold math
\begin{document}
% You should use BibTeX and apsrev.bst for references
\bibliographystyle{apsrev}

%\preprint{}

\title{Effect of two gaps on the flux lattice internal field distribution: evidence of two length scales from $\mu$SR in Mg$_{1-x}$Al$_{x}$B$_2$.}
% Optional argument for running titles on pages
%\title[]{}


% other informatio
\author{S.~Serventi} 
\author{G.~Allodi} 
\author{R.~De Renzi} 
\author{G.~Guidi} 
\author{L.~Roman\`o}
%\thanks{}
\affiliation{Dipartimento di Fisica e Unit\`a INFM di Parma, I 43100 Parma, Italy}
\author{P.~Manfrinetti} 
\author{A.~Palenzona}
%\thanks{}
\affiliation{Dipartimento di Chimica Industriale e Unit\`a INFM di Genova, I 16146 Genova, Italy}
\author{Ch.~Niedermayer}
\affiliation{Lab. for Neutron Scattering, ETH-Zurich and
Paul Scherrer Institut, CH-5232 Villigen PSI, Switzerland}
\author{A.~Amato} 
\author{Ch.~Baines}
%\thanks{}
\affiliation{Lab. for Muon Spin Spectroscopy, CH-5232 Villigen PSI, Switzerland}

\date{\today}

\begin{abstract}
We have measured the transverse field muon spin precession in the flux lattice (FL) state of the two gap superconductor MgB$_2$ and of the electron doped compounds Mg$_{1-x}$Al$_{x}$B$_2$ in magnetic fields up to 2.8T. We show the effect of the two gaps on the internal field distribution in the FL, from which we determine two coherence length parameters and the doping dependence of the London penetration depth. This is an independent determination of the complex vortex structure already suggested by the STM observation of large vortices in a MgB$_2$ single crystal. Our data  agrees quantitatively with STM and we thus validate a new phenomenological model for the internal fields.

\end{abstract}
%\keywords{}

\maketitle


%\label{}
%\subsection{}
%\subsubsection{}

% If in two-column mode, this environment will change to single-column
% format so that long equations can be displayed. Use
%\begin{widetext}
%\end{widetext}

The electronic and superconducting\cite{Nagamatsu:2001} properties of MgB$_2$ are determined by the presence of two types of carriers (i.e. bands crossed by the Fermi surface), electrons in the $\pi$ bands characteristic of all graphite-like materials, and holes in the nearly two dimensional $\sigma$ bands. There is a large consensus on the origin of the large superconducting $T_c=39$ K being due to the high energy phonon mode and to the large density of states connected with the narrow $\sigma$ bands\cite{Kortus:2001}. However also $\pi$ electrons contribute essentially to superconductivity and, in zero magnetic field, the transition temperatures of the two fluids appear to coincide\cite{Gonnelli:2002}.

Magnetic properties (e.g.~field penetration) of a two gap superconductor are not fully established yet. In particular the investigation of the field distribution in the flux lattice (FL), a specialty of $\mu$SR, is still at an early stage in the presence of two gaps, whereas it is theoretically and experimentally very well developed for single gap materials\cite{Brandt:1988,Barford:1988,Hao:1991,Pumpin:1990,Herlach:1990,Sonier:2000}. MgB$_2$ is the ideal simple material for such an investigation. 

Efforts have been devoted to the intermediate task of calculating\cite{Golubov:2002} the London penetration depth $\lambda$ within the Eliashberg scheme. Another notable handle toward internal field distribution was offered by STM single crystal measurements\cite{Eskildsen:2002} with $\bm{H}\parallel c$. This imaging technique directly determines a large core dimension $\xi_v=50$ nm, at applied fields surprisingly exceeding the corresponding $\mu_0H_{c2}=0.13$ T. Calculations by Usadel equations\cite{Koshelev:2003} justify this observation introducing two length scales, $\xi_\sigma$ and $\xi_\pi$, within a moderately dirty picture of MgB$_2$. Additional evidence for a direct influence of two distinct densities of supercarriers on the flux lattice and its  field dependence was gathered from a small angle neutron scattering (SANS) experiment\cite{Cubitt:2003} which detects a crossover field $\mu_0H=0.5$ T connected with field induced depletion of $\pi$ pairs.

The well characterized alloy Mg$_{1-x}$Al$_{x}$B$_2$ provides an additional experimental handle, since the doping effect of the Al$^{3+}$-Mg$^{2+}$ substitution as well as its influence on the phonon spectrum, both accurately described by band calculations\cite{Liu:2001,Satta:2001,Boeri:2002} lead to the vanishing of $T_c$ for $x\approx 0.45$. The key feature is that increasing concentrations of Al  produce a progressive filling of the $\sigma$ band, hence the stoichiometry offers independent experimental control over the two superconducting fluids. 

Early $\mu$SR results\cite{Niedermayer:2002} on the pure sample revealed a large variation of the second moment $\sigma_\mu^2$ of the muon line with external field, up to 0.6 T, attributing the dependence to low field pinning effects. In extreme type II materials $\sigma_\mu$ would be proportional to the inverse squared London penetration depth. Subsequent studies \cite{Serventi:2003,Papagelis:2003} detected a reduction of $\sigma_\mu$ in Al and C substituted powders, consistent with filling of the $\sigma$ band. Recent measurements \cite{Ohishi:2003} extended the field range to 5 T and attributed $\sigma_\mu(H)$ to a field dependence of the coherence length. 
We present here extended data on Mg$_{1-x}$Al$_{x}$B$_2$, $0\le x\le 0.35$ and offer a novel interpretation of the field dependence.

Powder samples of Mg$_{1-x}$Al$_{x}$B$_2$ were prepared according to standard methods \cite{Putti:2003}. The experiments were performed on loose powders wrapped in thin Al foil as well as on a high density (90\% of the bulk value) disk pellet of pure MgB$_2$. 
The experiments were performed on the GPS and LTF spectrometers of the S$\mu$S, Paul Scherrer Institut, Villigen, Switzerland.

The samples were cooled in a magnetic field $\mu_0 \bm {H}$ establishing a flux lattice (FL) below $T_c$. The muon spin precession signal was collected in two detectors, 1 and 2, at the opposite ends of an axis perpendicular to $\bm H$, in a transverse field (TF) $\mu$SR setup. The asymmetry in the muon decay is obtained from the count rates $N_i$ in the two detectors as ${\cal A}(t) = (N_1(t)-\alpha N_2(t))/(N_1(t)+\alpha N_2(t))$; the effective relative count efficiency $\alpha$ is calibrated in the normal metallic phase. 
 \begin{figure}
\includegraphics[width=0.4\textwidth]{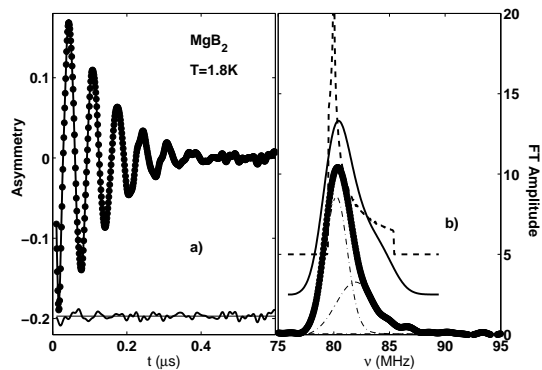}%
 \caption{TF  $\mu$SR $x=0$ data in 0.6 T: a) rotating frame asymmetry ($\nu=65$ MHz, symbols), best fit (solid line, Eq.~\ref{eq:asymmetry}) and fit residues (solid line, displaced for clarity); b) FFT of the muon asymmetry (symbols) and of the two components of its best fit (dash-dot lines, their sum falls under the symbols); also shown, but displaced vertically for clarity, are the FL lineshape $p(\gamma B)$ (dashed line, $\lambda=52$ nm, $\xi=23$ nm) and the Gaussian-convoluted FL lineshape (solid line, same $\lambda$, $\xi$, Gaussian half width $\Delta w=1$ MHz).  }
 \label{fig:1}
 \end{figure}
Figure \ref{fig:1} shows a typical precession asymmetry at $T=1.8$ K, transformed, for visualization purposes, to a rotating frame at a frequency $\omega=\omega_L-\delta\omega$, close to the Larmor frequency $\omega_L=2\pi\gamma B$, where $\gamma=135.5$ MHz/T is the muon magnetogyric ratio and $B$ the modulus of the local field. The damping is due to the FL field distribution, since muons stopping close to the vortex cores precess faster than those midway between two vortices.
The standard practice\cite{Niedermayer:2002,Sonier:2000} for polycrystal data is to fit asymmetries to $A e^{-t^2\sigma_\mu^2/2} \cos(\omega t+\phi)$ and to assume the second moment $\overline{\Delta B^2}=\overline{B^2}-\overline{B}^2$ of the FL field distribution proportional to the Gaussian second moment, $\sigma_\mu^2\approx(2\pi\gamma)^2 \overline{\Delta B^2}$. 

The one-gap calculation\cite{Brandt:1988} of the FL field distribution $p(B)$ is given by the dashed line in Fig.~\ref{fig:1}b (where $\nu=\gamma B$), with its characteristic peak at the saddle point between two adjacent fluxons.  Also shown is the convolution of $p(B)$ with a Gaussian, to empirically account for FL disorder introduced by vortex pinning and for the distribution of demagnetizing fields in the polycrystal grains. The curves are obtained with reasonable, although non optimized values of the coherence length,  $\xi$, and of the London penetration depth,  $\lambda$. 

The fast Fourier transform (FFT) of the precession asymmetry (Fig.\ref{fig:1}b, filled symbols) reflects\cite{Sonier:2000} directly $p(B)$ and displays a very similar asymmetric shape. A two Gaussian model 
\begin{equation} 
A_{fit}(t)=A_1 e^{-t^2\sigma_{\mu 1}^2/2} \cos(\omega_1 t+\phi)+A_2 e^{-t^2\sigma_{\mu 2}^2/2} \cos(\omega_2 t+\phi);
\label{eq:asymmetry}
\end{equation}
improves significantly a large fraction of our fits, in particular high statistics ones. The FFT of the two terms are shown by dashed-dotted lines in Fig.~\ref{fig:1}b. The solid line in Fig.~\ref{fig:1}a actually shows the best fit to Eq.~\ref{eq:asymmetry}, displaying negligible fit residuals, ${\cal A}-A_{fit}$. The second moment of the composite lineshape may be calculated as
\begin{equation} 
\sigma_\mu^2 = a_1 \sigma_{\mu 1}^2 + a_2 \sigma_{\mu 2}^2 + (a_1-a_1^2)\omega_1^2 + (a_2-a_2^2)\omega_2^2 - 2 a_1 a_2 \omega_1 \omega_2
\label{eq:sigmafl}
\end{equation}
where $a_i=A_i/(A_1+A_2)$ are weights and $\omega_i$ the first moments of the two components. 

The experimental values of $\sigma_\mu$ thus obtained may be compared with the FL prediction \cite{Brandt:1988} for $\overline{\Delta B^2}$, which is derived from the expression of the local field intensity at point $\bm{r}$ 
\begin{eqnarray} 
B(\bm{r}) & = &\mu_0 H \sum_{\bm{q}} \frac{ e^{-q^2\xi^2/2(1-h)}} {1+q^2\lambda^2/(1-h)}\, e^{i\bm{q}\cdot\bm{r}}
\label{eq:field}
\\ 
\overline{\Delta B^2} & = & \mu_0^2 H^2 \sum_{\bm{q}}\left [ \frac{e^{-q^2\xi^2/2(1-h)}} {1+q^2\lambda^2/(1-h)} \right]^2,
\label{eq:secondmoment}
\end{eqnarray}
where $\xi$ and $\lambda$ are divided by $\sqrt{1-h}$, with $h=H/H_{c2}$, to account\cite{Sonier:2000} for the field dependence of the Ginzburg-Landau order parameter.
The distribution $p(B)$ in Fig.~\ref{fig:1} is directly derived from Eq.~\ref{eq:field}, whose dependence on $H$ is essentially due to the reciprocal FL vector $\bm q$, since the FL lattice cell area is equal to $\phi_0/\mu_0H$, where $\phi_0$ is the flux quantum, hence $q\propto \sqrt{H}$.

In the extreme type II case ($\lambda\gg\xi$), since in practice $H\ll H_{c2}$, the second moment\cite{Brandt:1988,Barford:1988}  does not depend on either $H$ or $\xi$. In these conditions the spatial variation of $B(r)/\mu_0H$ is governed only by $\lambda$, and one has $\sigma_\mu(\mu\mbox{s}^{-1})=0.10734\lambda^{-2}(\mu\mbox{m}^{-2})$. However, when $\xi$ is just moderately smaller than $\lambda$, the finite size of the flux core produces an excluded volume effect \cite{Hillier:1997} and the resulting decrease of $\sigma_\mu(H)$ vs. $H$ is controlled by $\xi$. The cutoff function $e^{-q^2\xi^2/2}$ is a standard \cite{Brandt:1988} approximation to the solution of Ginzburg Landau equations \cite{Hao:1991}, which involves a modified Bessel function cutoff. An alternative $e^{-\sqrt{2}q\xi}$ approximation was also suggested\cite{Yaouanc:1997}. 

 \begin{figure}
\includegraphics[width=0.4\textwidth]{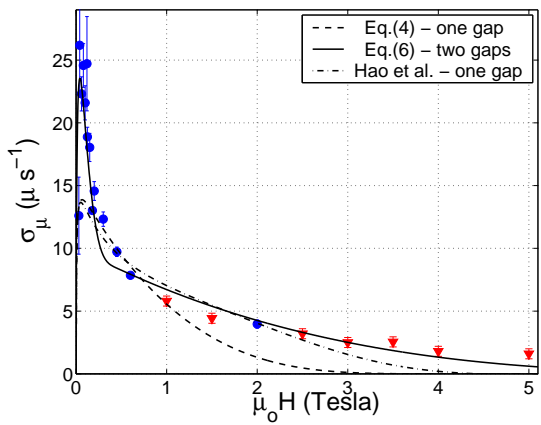}%
 \caption{$\mu$SR second moments in MgB$_2$ from Eq.~\ref{eq:sigmafl} ($\bullet$), including the $\mu_0 H>0.6$ T data from Ref.~\onlinecite{Ohishi:2003} ($\blacktriangledown$, single Gaussian fit); the dashed, dahed-dotted lines are one-gap best fits (see text); the solid line is the two-gap best fit of Eq.~\ref{eq:twogapssecond} }
 \label{fig:2}
 \end{figure}

Figure \ref{fig:2} shows the values of $\sigma_\mu$ obtained by Eq.~\ref{eq:sigmafl} from $T=1.8$ K measurements as a function of $\mu_0 H$, for pure MgB$_2$.  The high field single Gaussian widths of Ref.~\onlinecite{Ohishi:2003} are also included. Their field dependence agrees well with ours, despite the different analysis. Incidentally the difference between values derived from Eq.~\ref{eq:sigmafl} and single Gaussian widths is negligible also for our lower statistics low field data on pure MgB$_2$.

The dashed line shows the best fit to  Eq.~\ref{eq:secondmoment} ($\chi_r^2=17.5$ per degree of freedom, $\lambda=71$ nm, $\xi=9$ nm), which is poor and fails to reproduce the clear experimental change of slope around 0.5 T (very close to the crossover field  detected by SANS\cite{Cubitt:2003}). The dashed-dotted line shows the effect of a different cut-off function \cite{Hao:1991}, nearly equivalent to its approximation\cite{Yaouanc:1997},  which improves the fit ($\chi_r^2=X$), but is still poor both at low and at high fields.  We suggest that the inadequacy of Eq.~\ref{eq:secondmoment} is due to the two distinct densities of pairs evidenced by SANS.

A detailed calculation of $p(B)$ like those of Refs.~\onlinecite{Brandt:1988,Barford:1988,Hao:1991} has not been worked out to our knowledge for a two-gap superconductor. However, the extension of Ginzburg-Landau equations to two order parameters\cite{Golubov:2002,Koshelev:2003} indicates that $p(B)$ ought to be described in terms of a penetration depth $\lambda$ and two further length scales, a smaller $\xi_\sigma$ and a larger $\xi_\pi$. Each of these parameters corresponds to the square root of a diffusion coefficient\cite{Koshelev:2003} (the coherence length, for a clean, single gap system). One may therefore expect a change of slope in $\sigma_\mu(H)$ like that observed, when $H$ exceeds $H_{\pi}=\phi_0/2\pi \mu_0 \xi_\pi^2$. 

We therefore propose a simple extension of Eq.~\ref{eq:field}, by noting\cite{Sonier:2000} that, for intermediate fields, $B(\bm{r})\propto\lambda^{-2}$ is proportional to the superconducting carrier density $n$. Hence it is reasonable to assume 
\begin{eqnarray} 
B(\bm{r}) & = & \mu_0 H \sum_{\bm{q}}  \left[w_\sigma \frac{e^{-q^2\xi_\sigma^2/2(1-h_{\sigma})}} {1+q^2\lambda^2/(1-h_{\sigma})} \right. \nonumber \\   & + & \left. (1-w_\sigma) \frac{e^{-q^2\xi_\pi^2/2(1-h_{\pi})}} {1+q^2\lambda^2/(1-h_{\pi})}\right] \, e^{i\bm{q}\cdot\bm{r}}
\label{eq:twogapsfield}
\end{eqnarray}
where the weight $w_\sigma=n_\sigma/(n_\sigma+n_\pi)$, is proportional to the densities of superconducting carriers $n_{\sigma}$, $h_{\sigma,\pi}=H/H_{\sigma,\pi}$, and $H_{\sigma,\pi}=\phi_0/2\pi\mu_0 \xi_{\sigma,\pi}^2$. This expression leads to a new second moment expression

\begin{eqnarray} 
\overline{\Delta B^2} & = & \mu_0^2 H^2\sum_{\bm{q}} \left[ w_\sigma \frac{e^{-q^2\xi_\sigma^2/2(1-h_{\sigma})}} {1+q^2\lambda^2/(1-h_{\sigma})} \right. \nonumber \\ && \left. +  (1-w_\sigma) \frac{e^{-q^2\xi_\pi^2/2(1-h_{\pi})}} {1+q^2\lambda^2/(1-h_{\pi})} \right]^2
\label{eq:twogapssecond}
\end{eqnarray}
We notice that Eq.~\ref{eq:twogapssecond} does not follow the same assumption of previous works\cite{Niedermayer:2002,Papagelis:2003}, where the square root second moment $\sigma_\mu\propto\lambda^{-2}$ was taken as a {\em linear combination} of two terms. A linear $\lambda^{-2}=\omega_{p,\sigma}^2w^\prime_\sigma + \omega_{p,\pi}^2w^\prime_\pi$  is indeed that provided by two-gap Eliashberg theory\cite{Golubov:2002} in terms of the {\em superfluid plasma frequencies} $\omega_{p,\alpha}\sqrt{w_\alpha}$. However its direct proportionality with $\sigma_\mu$ was just assumed in analogy with the one-gap case, and this may well be unjustified in the multigap superconductor.

Figure \ref{fig:2} also shows the best fit to Eq.~\ref{eq:twogapssecond} (solid line, $\chi_r^2=5.8$, $\lambda=49(4)$ nm, $w_\sigma= 0.24(4)$ nm, $\xi_\sigma=5.1(2)$ nm, $\xi_\pi=23(1)$ nm) which features a change of slope around 0.5 T and follows much better the data to high fields, as the improved $\chi_r^2$ testifies. No futher appreciable improvement is obtained by changing the shape of the cut-off.

\begin{figure}
\includegraphics[width=0.4\textwidth]{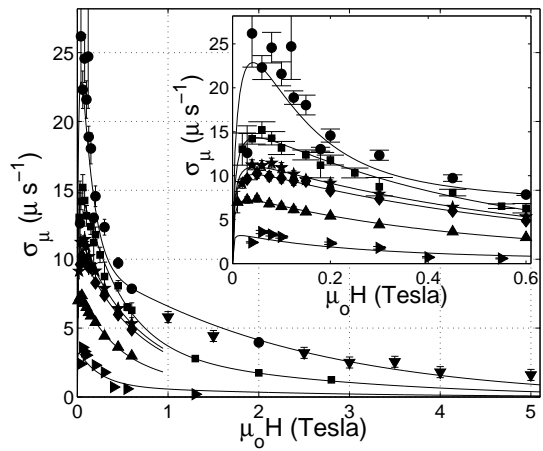}%
 \caption{$\mu$SR second moments in Mg$_{1-x}$Al$_x$B$_2$ from Eq.~\ref{eq:secondmoment} ($\bullet,\blacksquare, \bigstar, \blacklozenge, \blacktriangle,  \blacktriangleright$  respectively for $x=0,0.1,0.15,0.2,0.25,0.3$, including data from Ref.~\onlinecite{Ohishi:2003} ($\blacktriangledown,\, x=0$, single Gaussian fit); the solid line is the global two-gap fit (see text) of Eq.~\ref{eq:twogapssecond}. Inset: low field data.}
 \label{fig:3}
 \end{figure}

Our Mg$_{1-x}$Al$_x$B$_2$ data are shown in Fig.~\ref{fig:3} . It is apparent that increasing values of $x$ lead to a reduction of $\sigma_\mu$, i.e. to an increase\cite{Serventi:2003,Papagelis:2003} of the penetration depth $\lambda(x)$. This is not surprising since $\lambda^{-2}$ is proportional to an {\em effective} superconducting carrier density $n$, in analogy with the simple London model.

Since the most relevant high field data are missing for some  Al concentrations, a fully unconstrained fit of all four parameter can be performed only for three samples ($x=0,0.1,0.3$). In these fits $\xi_\sigma$ and $w_\sigma$ do not vary within errorbars. This is expected, since {\em i)} the $\sigma$ band is insensitive to impurity scattering and {\em ii)} $n_\pi$ must be a function of $n_\sigma$ ($\pi$ carriers on their own do not superconduct). Hence in first approximation $\xi_\sigma$ and $w_\sigma$ may be safely assumed independent of $x$.

The best fit shown by solid lines in Fig.~\ref{fig:3} corresponds to the following strategy: we performed a global best fit to Eq.~\ref{eq:twogapssecond} of the $x=0,0.1,0.3$ data, for which high field data are available. Thus we optimized $\lambda(x),\xi_\pi(x)$ and common {\em average} values of the other two parameters ($w_\sigma=0.29(4),\xi_\sigma=5.0(4)$ nm). Each of the $x=0.15,0.2,0.25$ data set was then fitted independently with the same fixed values of $w_\sigma,\xi_{\sigma}$, varying only $\lambda$ and $\xi_\pi$.

\begin{figure}
\includegraphics[width=0.47\textwidth]{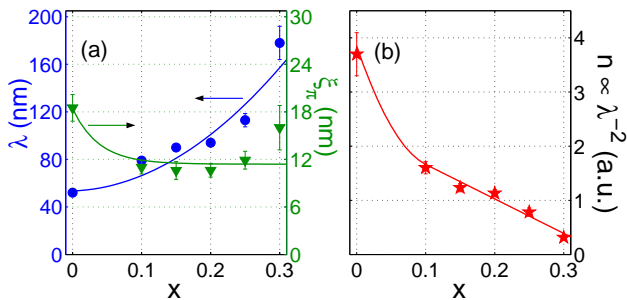}%
 \caption{ Mg$_{1-x}$Al$_x$B$_2$: a) penetration depth $\lambda$ ($\bullet$) and length $\xi_\pi$ ($\blacktriangledown$) vs. $x$ from the global best fit of Eq.~\ref{eq:twogapssecond}; b) effective supercarrier density $n\propto\lambda^{-2}$. Lines are guides to the eye.}
 \label{fig:4}
 \end{figure}
Our global fit procedure ($\chi^2_r=1.8$) determines the values of $\lambda$ and  $\xi_\pi$ shown in Fig.~\ref{fig:4}a, where $\lambda$ increases and $\xi_\pi$ decrease smoothly with $x$, as expected. The reduction of the effective length $\xi_\pi$ with doping indicates that the presence of Al reduces the $\pi$ carrier mean free path. 

Not surprisingly the $x=0$ value of $\lambda$ is smaller than previous determinations\cite{Niedermayer:2002,Ohishi:2003}, since this parameters describes the field profile around a vortex and we are now fitting a more complex fluxon structure. Fig.~\ref{fig:4}b plots the {\em effective } $n$~\-vs.~\-$x$ which displays a dramatic drop, extrapolating to zero already at $x=0.35$ (a sample where we failed to detect any FL). Note that our data represent a powder average, which masks, e.g.~\-$\sigma$ band effective mass anisotropy. However, as long as the anisotropy $(m_{c}/m_{ab})^{1/2}$ does not reduce drastically with doping, the $\mu$SR lineshape should be mostly sensitive\footnote{see for instance Eq.~6 in Ref.\onlinecite{Aegerter:1998}} to  the component for $\bm{H}\parallel c$.

%Our low field second moments are slightly larger than those of Ref.~\onlinecite{Niedermayer:2002,Ohishi:2003}, a fact which is intrinsic of the sample, independent of the details of data analysis (we checked it on Ref.~\onlinecite{Niedermayer:2002} data).

Our interpretation can be reconciled with that of Ref.\cite{Ohishi:2003}, whose  MgB$_2$ data are included in our fit. The authors invoke a field dependence of the (single) coherence length, which may well actually correspond to a field dependent weight of two length scales. However our analysis supports explicitly the existence of two scales in the system, and it allows a quantitative comparison with Usadel calculations, as well as with the STM data\cite{Eskildsen:2002}. As a matter of fact the latter imply a complex vortex structure, since they  determines a large $\xi_v=49.6(9)$ nm value, although their measured gap $\Delta_\sigma$ yields $\xi_\sigma=13$ nm and the observed upper critical field is consistent with a $\xi\ll \xi_v$.

The Usadel calculations\cite{Koshelev:2003} do justify this observation. They predict an apparent vortex length scale $\xi_v=2.7\xi_\pi+0.3\xi_\sigma$, which corresponds to 52(4) nm if we identify their length scales with our global fit parameters (the large uncertainty accounts for the variation of these parameters with cut-off and fitting strategy). The agreement is very good and it provides quantitative support to our model.

In conclusion we have shown that two length scales govern the field distribution inside a two gap superconductor, we have measured the average supercarrier density as a function of the $\sigma$ band filling and found remarkable agreement with Ginzburg-Landau calculations and STM determinations of the vortex dimension.
%\begin{acknowledgments}
This work was financially supported by the Federal Office for Education and Science, Bern, Switzerland. RDR and AP acknowledge the support of INFM  grant PRA-UMBRA. 
\bibliography{MgB2}

\begin{thebibliography}{24}
\expandafter\ifx\csname natexlab\endcsname\relax\def\natexlab#1{#1}\fi
\expandafter\ifx\csname bibnamefont\endcsname\relax
  \def\bibnamefont#1{#1}\fi
\expandafter\ifx\csname bibfnamefont\endcsname\relax
  \def\bibfnamefont#1{#1}\fi
\expandafter\ifx\csname citenamefont\endcsname\relax
  \def\citenamefont#1{#1}\fi
\expandafter\ifx\csname url\endcsname\relax
  \def\url#1{\texttt{#1}}\fi
\expandafter\ifx\csname urlprefix\endcsname\relax\def\urlprefix{URL }\fi
\providecommand{\bibinfo}[2]{#2}
\providecommand{\eprint}[2][]{\url{#2}}

\bibitem[{\citenamefont{Nagamatsu et~al.}(2001)\citenamefont{Nagamatsu,
  Nakagawa, Muranaka, Zenitani, and Akimitsu}}]{Nagamatsu:2001}
\bibinfo{author}{\bibfnamefont{J.}~\bibnamefont{Nagamatsu}},
  \bibinfo{author}{\bibfnamefont{N.}~\bibnamefont{Nakagawa}},
  \bibinfo{author}{\bibfnamefont{T.}~\bibnamefont{Muranaka}},
  \bibinfo{author}{\bibfnamefont{Y.}~\bibnamefont{Zenitani}}, \bibnamefont{and}
  \bibinfo{author}{\bibfnamefont{J.}~\bibnamefont{Akimitsu}},
  \bibinfo{journal}{Nature} \textbf{\bibinfo{volume}{410}}, \bibinfo{pages}{63}
  (\bibinfo{year}{2001}).

\bibitem[{\citenamefont{Kortus et~al.}(2001)\citenamefont{Kortus, Mazin,
  Belashchenko, Antropov, and Boyer}}]{Kortus:2001}
\bibinfo{author}{\bibfnamefont{J.}~\bibnamefont{Kortus}},
  \bibinfo{author}{\bibfnamefont{I.}~\bibnamefont{Mazin}},
  \bibinfo{author}{\bibfnamefont{K.}~\bibnamefont{Belashchenko}},
  \bibinfo{author}{\bibfnamefont{V.}~\bibnamefont{Antropov}}, \bibnamefont{and}
  \bibinfo{author}{\bibfnamefont{L.}~\bibnamefont{Boyer}},
  \bibinfo{journal}{Phys.~Rev.Lett.} \textbf{\bibinfo{volume}{86}},
  \bibinfo{pages}{4656} (\bibinfo{year}{2001}).

\bibitem[{\citenamefont{Gonnelli et~al.}(2002)\citenamefont{Gonnelli, Daghero,
  Ummarino, Stepanov, Jun, Kazakov, and Karpinski}}]{Gonnelli:2002}
\bibinfo{author}{\bibfnamefont{R.}~\bibnamefont{Gonnelli}},
  \bibinfo{author}{\bibfnamefont{D.}~\bibnamefont{Daghero}},
  \bibinfo{author}{\bibfnamefont{G.}~\bibnamefont{Ummarino}},
  \bibinfo{author}{\bibfnamefont{V.}~\bibnamefont{Stepanov}},
  \bibinfo{author}{\bibfnamefont{J.}~\bibnamefont{Jun}},
  \bibinfo{author}{\bibfnamefont{S.}~\bibnamefont{Kazakov}}, \bibnamefont{and}
  \bibinfo{author}{\bibfnamefont{J.}~\bibnamefont{Karpinski}},
  \bibinfo{journal}{Phys. Rev. Lett.} \textbf{\bibinfo{volume}{89}},
  \bibinfo{pages}{247004} (\bibinfo{year}{2002}).

\bibitem[{\citenamefont{Brandt}(1988)}]{Brandt:1988}
\bibinfo{author}{\bibfnamefont{E.~H.} \bibnamefont{Brandt}},
  \bibinfo{journal}{Phys. Rev.} \textbf{\bibinfo{volume}{B37}},
  \bibinfo{pages}{R2349} (\bibinfo{year}{1988}).

\bibitem[{\citenamefont{Barford and Gunn}(1988)}]{Barford:1988}
\bibinfo{author}{\bibfnamefont{W.}~\bibnamefont{Barford}} \bibnamefont{and}
  \bibinfo{author}{\bibfnamefont{J.}~\bibnamefont{Gunn}},
  \bibinfo{journal}{Phys.~C.} \textbf{\bibinfo{volume}{156}},
  \bibinfo{pages}{515} (\bibinfo{year}{1988}).

\bibitem[{\citenamefont{Hao et~al.}(1991)\citenamefont{Hao, Clem, Civale,
  Malozemoff, and Holtzberg}}]{Hao:1991}
\bibinfo{author}{\bibfnamefont{Z.}~\bibnamefont{Hao}},
  \bibinfo{author}{\bibfnamefont{J.}~\bibnamefont{Clem}},
  \bibinfo{author}{\bibfnamefont{M.~M.~L.} \bibnamefont{Civale}},
  \bibinfo{author}{\bibfnamefont{A.}~\bibnamefont{Malozemoff}},
  \bibnamefont{and}
  \bibinfo{author}{\bibfnamefont{F.}~\bibnamefont{Holtzberg}},
  \bibinfo{journal}{Phys. Rev.} \textbf{\bibinfo{volume}{B43}},
  \bibinfo{pages}{2844} (\bibinfo{year}{1991}).

\bibitem[{\citenamefont{Pumpin et~al.}(1990)\citenamefont{Pumpin, Keller,
  Kundig, Odermatt, Savic, Schneider, Simmler, Zimmermann, Kaldis, Rusiecki
  et~al.}}]{Pumpin:1990}
\bibinfo{author}{\bibfnamefont{B.}~\bibnamefont{Pumpin}},
  \bibinfo{author}{\bibfnamefont{H.}~\bibnamefont{Keller}},
  \bibinfo{author}{\bibfnamefont{W.}~\bibnamefont{Kundig}},
  \bibinfo{author}{\bibfnamefont{W.}~\bibnamefont{Odermatt}},
  \bibinfo{author}{\bibfnamefont{I.}~\bibnamefont{Savic}},
  \bibinfo{author}{\bibfnamefont{J.}~\bibnamefont{Schneider}},
  \bibinfo{author}{\bibfnamefont{H.}~\bibnamefont{Simmler}},
  \bibinfo{author}{\bibfnamefont{P.}~\bibnamefont{Zimmermann}},
  \bibinfo{author}{\bibfnamefont{E.}~\bibnamefont{Kaldis}},
  \bibinfo{author}{\bibfnamefont{S.}~\bibnamefont{Rusiecki}},
  \bibnamefont{et~al.}, \bibinfo{journal}{Phys. Rev.}
  \textbf{\bibinfo{volume}{B42}}, \bibinfo{pages}{8019} (\bibinfo{year}{1990}).

\bibitem[{\citenamefont{Herlach et~al.}(1990)\citenamefont{Herlach, Majer,
  Major, Rosenkranz, Schmolz, Schwarz, Seeger, Templ, Brandt, Essmann
  et~al.}}]{Herlach:1990}
\bibinfo{author}{\bibfnamefont{D.}~\bibnamefont{Herlach}},
  \bibinfo{author}{\bibfnamefont{G.}~\bibnamefont{Majer}},
  \bibinfo{author}{\bibfnamefont{J.}~\bibnamefont{Major}},
  \bibinfo{author}{\bibfnamefont{J.}~\bibnamefont{Rosenkranz}},
  \bibinfo{author}{\bibfnamefont{M.}~\bibnamefont{Schmolz}},
  \bibinfo{author}{\bibfnamefont{W.}~\bibnamefont{Schwarz}},
  \bibinfo{author}{\bibfnamefont{A.}~\bibnamefont{Seeger}},
  \bibinfo{author}{\bibfnamefont{W.}~\bibnamefont{Templ}},
  \bibinfo{author}{\bibfnamefont{E.}~\bibnamefont{Brandt}},
  \bibinfo{author}{\bibfnamefont{U.}~\bibnamefont{Essmann}},
  \bibnamefont{et~al.}, \bibinfo{journal}{Hyperf.~Interactions}
  \textbf{\bibinfo{volume}{63}}, \bibinfo{pages}{41} (\bibinfo{year}{1990}).

\bibitem[{\citenamefont{Sonier et~al.}(2000)\citenamefont{Sonier, Brewer, and
  Kiefl}}]{Sonier:2000}
\bibinfo{author}{\bibfnamefont{J.}~\bibnamefont{Sonier}},
  \bibinfo{author}{\bibfnamefont{J.}~\bibnamefont{Brewer}}, \bibnamefont{and}
  \bibinfo{author}{\bibfnamefont{R.}~\bibnamefont{Kiefl}},
  \bibinfo{journal}{Rev. Mod. Phys.} \textbf{\bibinfo{volume}{72}},
  \bibinfo{pages}{769} (\bibinfo{year}{2000}).

\bibitem[{\citenamefont{Golubov et~al.}(2002)\citenamefont{Golubov, Brinkman,
  Dolgov, Kortus, and Jepsen}}]{Golubov:2002}
\bibinfo{author}{\bibfnamefont{A.}~\bibnamefont{Golubov}},
  \bibinfo{author}{\bibfnamefont{A.}~\bibnamefont{Brinkman}},
  \bibinfo{author}{\bibfnamefont{O.}~\bibnamefont{Dolgov}},
  \bibinfo{author}{\bibfnamefont{J.}~\bibnamefont{Kortus}}, \bibnamefont{and}
  \bibinfo{author}{\bibfnamefont{O.}~\bibnamefont{Jepsen}},
  \bibinfo{journal}{Phys.~Rev.} \textbf{\bibinfo{volume}{B66}},
  \bibinfo{pages}{054524} (\bibinfo{year}{2002}).

\bibitem[{\citenamefont{Eskildsen et~al.}(2002)\citenamefont{Eskildsen, Kugler,
  Tanaka, Jun, Kazakov, Karpinski, and Fischer}}]{Eskildsen:2002}
\bibinfo{author}{\bibfnamefont{M.}~\bibnamefont{Eskildsen}},
  \bibinfo{author}{\bibfnamefont{M.}~\bibnamefont{Kugler}},
  \bibinfo{author}{\bibfnamefont{S.}~\bibnamefont{Tanaka}},
  \bibinfo{author}{\bibfnamefont{J.}~\bibnamefont{Jun}},
  \bibinfo{author}{\bibfnamefont{S.}~\bibnamefont{Kazakov}},
  \bibinfo{author}{\bibfnamefont{J.}~\bibnamefont{Karpinski}},
  \bibnamefont{and} \bibinfo{author}{\bibfnamefont{O.}~\bibnamefont{Fischer}},
  \bibinfo{journal}{Phys.~Rev.~Lett.} \textbf{\bibinfo{volume}{89}},
  \bibinfo{pages}{187003} (\bibinfo{year}{2002}).

\bibitem[{\citenamefont{Koshelev and Golubov}(2003)}]{Koshelev:2003}
\bibinfo{author}{\bibfnamefont{A.}~\bibnamefont{Koshelev}} \bibnamefont{and}
  \bibinfo{author}{\bibfnamefont{A.}~\bibnamefont{Golubov}},
  \bibinfo{journal}{Phys.~Rev.~Lett.} \textbf{\bibinfo{volume}{90}},
  \bibinfo{pages}{177002} (\bibinfo{year}{2003}).

\bibitem[{\citenamefont{Cubitt et~al.}(2003)\citenamefont{Cubitt, Eskildsen,
  Dewhurst, Jun, Kazakov, and Karpinski}}]{Cubitt:2003}
\bibinfo{author}{\bibfnamefont{R.}~\bibnamefont{Cubitt}},
  \bibinfo{author}{\bibfnamefont{M.}~\bibnamefont{Eskildsen}},
  \bibinfo{author}{\bibfnamefont{C.}~\bibnamefont{Dewhurst}},
  \bibinfo{author}{\bibfnamefont{J.}~\bibnamefont{Jun}},
  \bibinfo{author}{\bibfnamefont{S.}~\bibnamefont{Kazakov}}, \bibnamefont{and}
  \bibinfo{author}{\bibfnamefont{J.}~\bibnamefont{Karpinski}},
  \bibinfo{journal}{Phys.~Rev.~Lett.} \textbf{\bibinfo{volume}{91}},
  \bibinfo{pages}{047002} (\bibinfo{year}{2003}).

\bibitem[{\citenamefont{Liu et~al.}(2001)\citenamefont{Liu, Mazin, and
  Kortus}}]{Liu:2001}
\bibinfo{author}{\bibfnamefont{A.}~\bibnamefont{Liu}},
  \bibinfo{author}{\bibfnamefont{I.}~\bibnamefont{Mazin}}, \bibnamefont{and}
  \bibinfo{author}{\bibfnamefont{J.}~\bibnamefont{Kortus}},
  \bibinfo{journal}{Phys.~Rev.~Lett.} \textbf{\bibinfo{volume}{87}},
  \bibinfo{pages}{087005} (\bibinfo{year}{2001}).

\bibitem[{\citenamefont{Satta et~al.}(2001)\citenamefont{Satta, Profeta,
  Bernardini, and Massidda}}]{Satta:2001}
\bibinfo{author}{\bibfnamefont{G.}~\bibnamefont{Satta}},
  \bibinfo{author}{\bibfnamefont{G.}~\bibnamefont{Profeta}},
  \bibinfo{author}{\bibfnamefont{F.}~\bibnamefont{Bernardini}},
  \bibnamefont{and} \bibinfo{author}{\bibfnamefont{A.~C.~S.}
  \bibnamefont{Massidda}}, \bibinfo{journal}{Phys.~Rev.}
  \textbf{\bibinfo{volume}{B64}}, \bibinfo{pages}{104507}
  (\bibinfo{year}{2001}).

\bibitem[{\citenamefont{Boeri et~al.}(2002)\citenamefont{Boeri, Bachelet,
  Cappelluti, and Pietronero}}]{Boeri:2002}
\bibinfo{author}{\bibfnamefont{L.}~\bibnamefont{Boeri}},
  \bibinfo{author}{\bibfnamefont{G.}~\bibnamefont{Bachelet}},
  \bibinfo{author}{\bibfnamefont{E.}~\bibnamefont{Cappelluti}},
  \bibnamefont{and}
  \bibinfo{author}{\bibfnamefont{L.}~\bibnamefont{Pietronero}},
  \bibinfo{journal}{Phys.~Rev.} \textbf{\bibinfo{volume}{B65}},
  \bibinfo{pages}{214501} (\bibinfo{year}{2002}).

\bibitem[{\citenamefont{Niedermayer et~al.}(2002)\citenamefont{Niedermayer,
  Bernhard, Holden, Kremer, and Ahn}}]{Niedermayer:2002}
\bibinfo{author}{\bibfnamefont{C.}~\bibnamefont{Niedermayer}},
  \bibinfo{author}{\bibfnamefont{C.}~\bibnamefont{Bernhard}},
  \bibinfo{author}{\bibfnamefont{T.}~\bibnamefont{Holden}},
  \bibinfo{author}{\bibfnamefont{R.~K.} \bibnamefont{Kremer}},
  \bibnamefont{and} \bibinfo{author}{\bibfnamefont{K.}~\bibnamefont{Ahn}},
  \bibinfo{journal}{Phys.~Rev.} \textbf{\bibinfo{volume}{B65}},
  \bibinfo{pages}{094512} (\bibinfo{year}{2002}).

\bibitem[{\citenamefont{Serventi et~al.}(2003)\citenamefont{Serventi, Allodi,
  Bucci, Renzi, Guidi, Palenzona, Manfrinetti, and Hillier}}]{Serventi:2003}
\bibinfo{author}{\bibfnamefont{S.}~\bibnamefont{Serventi}},
  \bibinfo{author}{\bibfnamefont{G.}~\bibnamefont{Allodi}},
  \bibinfo{author}{\bibfnamefont{C.}~\bibnamefont{Bucci}},
  \bibinfo{author}{\bibfnamefont{R.~D.} \bibnamefont{Renzi}},
  \bibinfo{author}{\bibfnamefont{G.}~\bibnamefont{Guidi}},
  \bibinfo{author}{\bibfnamefont{A.}~\bibnamefont{Palenzona}},
  \bibinfo{author}{\bibfnamefont{P.}~\bibnamefont{Manfrinetti}},
  \bibnamefont{and} \bibinfo{author}{\bibfnamefont{A.~D.}
  \bibnamefont{Hillier}}, \bibinfo{journal}{Physica} \textbf{\bibinfo{volume}{B
  326}}, \bibinfo{pages}{350} (\bibinfo{year}{2003}).

\bibitem[{\citenamefont{Papagelis et~al.}(2003)\citenamefont{Papagelis,
  Arvanitidis, Prassides, Schenck, Takenobu, and Iwasa}}]{Papagelis:2003}
\bibinfo{author}{\bibfnamefont{K.}~\bibnamefont{Papagelis}},
  \bibinfo{author}{\bibfnamefont{J.}~\bibnamefont{Arvanitidis}},
  \bibinfo{author}{\bibfnamefont{K.}~\bibnamefont{Prassides}},
  \bibinfo{author}{\bibfnamefont{A.}~\bibnamefont{Schenck}},
  \bibinfo{author}{\bibfnamefont{T.}~\bibnamefont{Takenobu}}, \bibnamefont{and}
  \bibinfo{author}{\bibfnamefont{Y.}~\bibnamefont{Iwasa}},
  \bibinfo{journal}{Europhys.~Lett.} \textbf{\bibinfo{volume}{61}},
  \bibinfo{pages}{254} (\bibinfo{year}{2003}).

\bibitem[{\citenamefont{Ohishi et~al.}(2003)\citenamefont{Ohishi, Muranaka,
  Akimitsu, Koda, Higemoto, and Kadono}}]{Ohishi:2003}
\bibinfo{author}{\bibfnamefont{K.}~\bibnamefont{Ohishi}},
  \bibinfo{author}{\bibfnamefont{T.}~\bibnamefont{Muranaka}},
  \bibinfo{author}{\bibfnamefont{J.}~\bibnamefont{Akimitsu}},
  \bibinfo{author}{\bibfnamefont{A.}~\bibnamefont{Koda}},
  \bibinfo{author}{\bibfnamefont{W.}~\bibnamefont{Higemoto}}, \bibnamefont{and}
  \bibinfo{author}{\bibfnamefont{R.}~\bibnamefont{Kadono}},
  \bibinfo{journal}{J.~Phys.~Soc.~Jpn.} \textbf{\bibinfo{volume}{72}},
  \bibinfo{pages}{29} (\bibinfo{year}{2003}).

\bibitem[{\citenamefont{Putti et~al.}(2003)\citenamefont{Putti, Affronte,
  Manfrinetti, and Palenzona}}]{Putti:2003}
\bibinfo{author}{\bibfnamefont{M.}~\bibnamefont{Putti}},
  \bibinfo{author}{\bibfnamefont{M.}~\bibnamefont{Affronte}},
  \bibinfo{author}{\bibfnamefont{P.}~\bibnamefont{Manfrinetti}},
  \bibnamefont{and}
  \bibinfo{author}{\bibfnamefont{A.}~\bibnamefont{Palenzona}},
  \bibinfo{journal}{Phys. Rev.} \textbf{\bibinfo{volume}{B68}},
  \bibinfo{pages}{094514} (\bibinfo{year}{2003}).

\bibitem[{\citenamefont{Hillier and Cywinski}(1997)}]{Hillier:1997}
\bibinfo{author}{\bibfnamefont{A.}~\bibnamefont{Hillier}} \bibnamefont{and}
  \bibinfo{author}{\bibfnamefont{R.}~\bibnamefont{Cywinski}},
  \bibinfo{journal}{Appl.~Mag.~Res..} \textbf{\bibinfo{volume}{13}},
  \bibinfo{pages}{95} (\bibinfo{year}{1997}).

\bibitem[{\citenamefont{Yaouanc et~al.}(1997)\citenamefont{Yaouanc, de~Reotier,
  and Brandt}}]{Yaouanc:1997}
\bibinfo{author}{\bibfnamefont{A.}~\bibnamefont{Yaouanc}},
  \bibinfo{author}{\bibfnamefont{P.~D.} \bibnamefont{de~Reotier}},
  \bibnamefont{and} \bibinfo{author}{\bibfnamefont{E.~H.}
  \bibnamefont{Brandt}}, \bibinfo{journal}{Phys.~Rev.~B}
  \textbf{\bibinfo{volume}{55}}, \bibinfo{pages}{11107} (\bibinfo{year}{1997}).

\bibitem[{\citenamefont{Aegerter et~al.}(1998)\citenamefont{Aegerter, Hofer,
  Savic, Keller, Lee, Ager, Lloyd, and Forgan}}]{Aegerter:1998}
\bibinfo{author}{\bibfnamefont{C.}~\bibnamefont{Aegerter}},
  \bibinfo{author}{\bibfnamefont{J.}~\bibnamefont{Hofer}},
  \bibinfo{author}{\bibfnamefont{I.}~\bibnamefont{Savic}},
  \bibinfo{author}{\bibfnamefont{H.}~\bibnamefont{Keller}},
  \bibinfo{author}{\bibfnamefont{S.}~\bibnamefont{Lee}},
  \bibinfo{author}{\bibfnamefont{C.}~\bibnamefont{Ager}},
  \bibinfo{author}{\bibfnamefont{S.}~\bibnamefont{Lloyd}}, \bibnamefont{and}
  \bibinfo{author}{\bibfnamefont{E.}~\bibnamefont{Forgan}},
  \bibinfo{journal}{Phys.~Rev.} \textbf{\bibinfo{volume}{B57}},
  \bibinfo{pages}{1253} (\bibinfo{year}{1998}).

\end{thebibliography}

\end{document}